\documentclass[pra,showpacs,twocolumn,superscriptaddress,longbibliography]{revtex4-2}

\usepackage{amsmath}
\usepackage{amssymb}
\usepackage{graphicx}
\usepackage{dcolumn}
\usepackage{bm}
\usepackage{amsfonts}
\usepackage{float}
\usepackage{color}
\usepackage{xcolor}
\usepackage{hyperref}

\def\be{\begin{equation}}
\def\ee{\end{equation}}

\begin{document}
\title{Observation of self-oscillating supersonic flow across an acoustic horizon in two dimensions}
\author{Hikaru Tamura}
\thanks{Present address: Institute for Molecular Science, Okazaki, Aichi, 444-8585, Japan}
\affiliation{Department of Physics and Astronomy, Purdue University, West Lafayette, IN 47907, USA}
\author{Sergei Khlebnikov}
\affiliation{Department of Physics and Astronomy, Purdue University, West Lafayette, IN 47907, USA}
\affiliation{Purdue Quantum Science and Engineering Institute, Purdue University, West Lafayette, IN 47907, USA}
\author{Cheng-An Chen}
\thanks{Current address: Atom Computing, Boulder, CO 80301, USA}
\affiliation{Department of Physics and Astronomy, Purdue University, West Lafayette, IN 47907, USA}
\author{Chen-Lung Hung}
\email{clhung@purdue.edu}
\affiliation{Department of Physics and Astronomy, Purdue University, West Lafayette, IN 47907, USA}
\affiliation{Purdue Quantum Science and Engineering Institute, Purdue University, West Lafayette, IN 47907, USA}
\date{\today}

\begin{abstract}
We report observation of self-oscillating supersonic flows in a two-dimensional atomic superfluid. By imposing a local particle sink with strong loss, we induce a convergent radial flow with a spatially bounded supersonic region, 
forming an acoustic analogue of a black-hole horizon and an inner horizon around the sink. The observed superflow appears to be modulated by quasi-periodic bursts of superluminal signals. We measure their frequencies and find agreement with numerical simulations of the frequencies of ring soliton oscillations within the black-hole horizon. The solitons seen in the simulations are emitted from the region between the two horizons in a process that we attribute to the Landau instability. The presented experiment demonstrates a new method for creating supersonic flows in atomic superfluids, which may find applications in quantum simulations of curved spacetime, supersonic turbulence, and self-oscillating dynamics in dissipative many-body systems.
\end{abstract}
\maketitle

According to Landau's criterion of superfluidity~\cite{landau1941theory}, a superfluid flowing past an obstacle becomes unstable with respect to production of excitations when the velocity exceeds a certain limit. For point-like obstacles in a weakly interacting Bose-Einstein condensate (BEC), phonon excitations dominate, and the critical velocity coincides with the speed of sound. When the obstacle size increases and becomes comparable to the healing length, much lower critical velocities are observed, which have been attributed to production of low energy vortex excitations~\cite{feynman1955chapter,neely2010observation,ramanathan2011superflow,desbuquois2012superfluid}. For a one-dimensional (1D) superflow, on the other hand, the critical velocity~\cite{Engels2007} has been found to depend on the obstacle height~\cite{hakim1997nonlinear,pavloff2002breakdown}, and it has been suggested that the Landau instability sets in when the local flow velocity exceeds the local sound speed~\cite{pavloff2002breakdown,finazzi2015instability}. 

More generally, even without an obstacle, one expects that the Landau instability plays a role as long as translational symmetry is broken. An intriguing, yet unexplored example is a convergent two-dimensional (2D) radial flow, where the flow rate grows with falling radius $r$. The flow could become unstable at a small enough radius, where the flow rate exceeds the local sound speed. In many related settings in 1D, Landau instability manifests itself through periodic emission of solitons~\cite{hakim1997nonlinear,leboeuf2001bose,pavloff2002breakdown,Engels2007,michel2015nonlinear,de2016time}. Here, we explore the stability of a 2D radial flow and report observation of quasi-periodic oscillations.

Intriguingly, spatially inhomogeneous flows with a subsonic-to-supersonic transition have been theorized \cite{unruh1981experimental,barcelo2011analogue} and broadly pursued (see~\cite{munoz2019observation,drori2019observation,euve2016observation,nguyen2015acoustic,weinfurtner2011measurement,philbin2008fiber} for examples) as simulators of an elusive phenomenon---Hawking radiation from a black hole horizon~\cite{hawking1975particle}. An acoustic black-hole (white-hole) horizon marks the transition of a subsonic flow to (from) a supersonic region that low-frequency sound waves cannot escape (re-enter). A bounded supersonic flow, like those in a penetrable barrier or in a convergent 2D flow, is enclosed by a pair of black-hole and white-hole (inner) horizons. In the presence of superluminal (faster than sound) short-wave excitations, a pair of acoustic horizons can act like mirrors that form a laser cavity, further amplifying the out-going Hawking radiation via stimulated emission~\cite{corley1999black}. More generally, Hawking radiation can also be stimulated from an initial seed produced by other mechanisms~\cite{weinfurtner2011measurement}. Recent discussions of this effect in 1D include~\cite{leonhardt2007black,finazzi2010black,coutant2010black,finazzi2015instability}.

In contrast to Hawking radiation, soliton and wave emissions due to a classical Landau instability do not require initial seeds. Testing the instability of supersonic flow within two horizons~\cite{finazzi2015instability,michel2015nonlinear,de2016time,de2021long} has so far remained an open experimental question. For instance, a recent experiment by the Technion group \cite{munoz2019observation} has generated acoustic horizons by sweeping a potential step along an elongated condensate. This method has led to a successful observation of Hawking radiation of phonons \cite{steinhauer2016observation,munoz2019observation} across a horizon that co-moves with the step potential. Phonons emitted following formation of an inner horizon, however, have been attributed not to the spontaneous Hawking process but to an amplification of the Cherenkov radiation from a moving obstacle \cite{tettamanti2016numerical,wang2017mechanism,kolobov2021observation}.

\begin{figure}[h!]
\centering
\includegraphics[width=1\columnwidth]{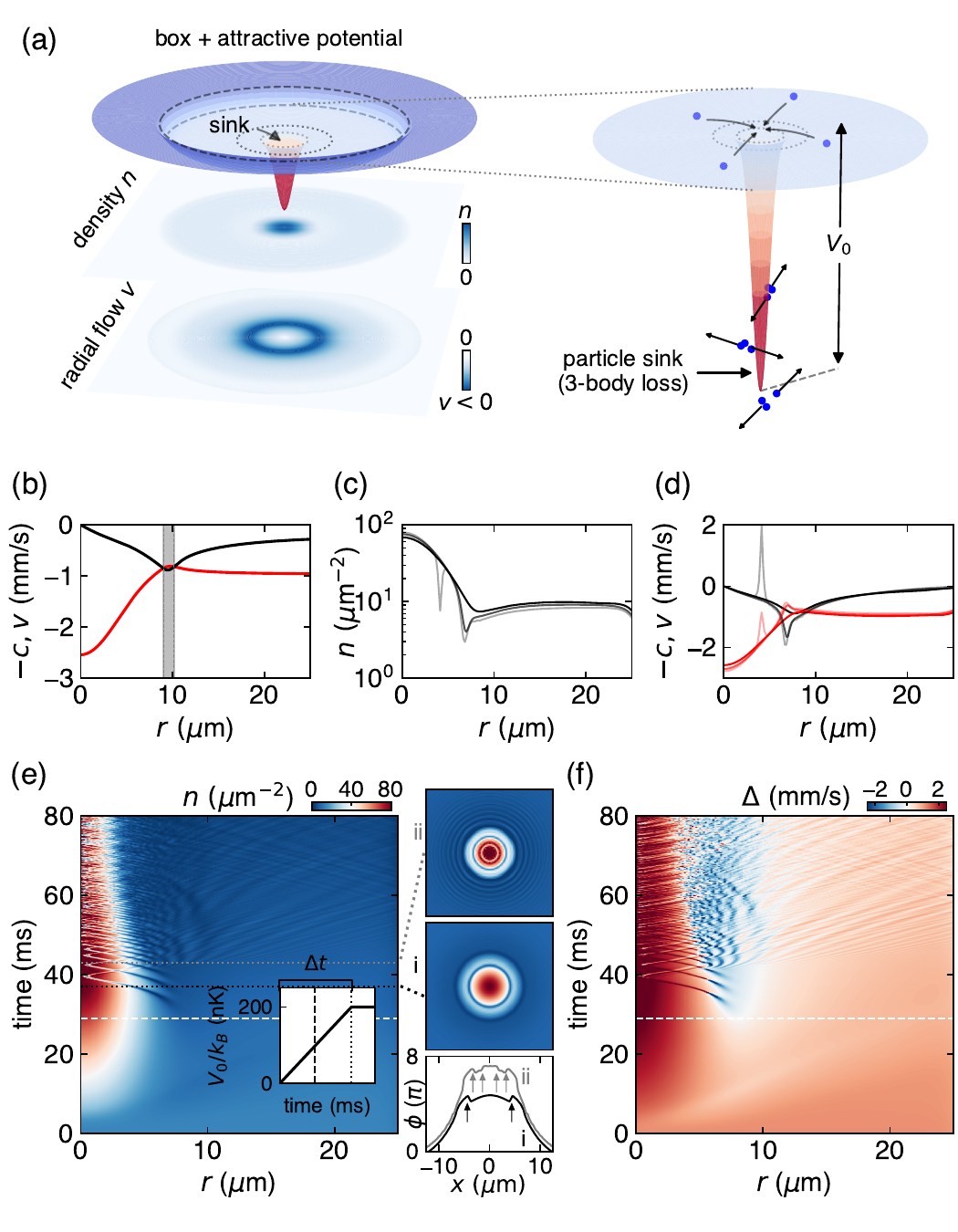}
\caption{Supersonic flow induced by a particle sink. (a) Schematics of a Gaussian attractive potential (depth $V_0$) imposing on a 2D atomic superfluid trapped in a circular box. The high density region with a large three-body recombination loss rate serves as a particle sink, inducing strong radial flow (velocity $v<0$). Dotted circles illustrate an acoustic black-hole horizon and an inner horizon. (b) Velocity $v$ (black) and local sound speed $c=\hbar\sqrt{ng}/m$ (red) versus radial position $r$ 
for the stationary solution of the GPE in an effectively infinite system at a critical depth $V_0=V_\mathrm{cr}$. Shaded region marks the supersonic flow $|v(r)|> c(r)$. For a finite system, time-dependent simulations suggest that ramping on a sink potential beyond the critical depth triggers soliton emission. (c) Density profiles at 0, 4, and 8~ms (dark to light gray curves) right after supersonic flow forms, obtained 
from a numerical solution of the time-dependent 2D GPE using a slow ramp of $V_0$ ($\Delta t=60$~ms as illustrated in (e) inset). (d) Sound (red) and flow velocity (gray) profiles corresponding to those in (c). (e-f) Full numerical time evolution of $n(r)$ and $\Delta(r)= c(r)+v(r)$, showing initial soliton emission near the critical depth (marked by dashed lines), radial oscillations of solitons, and multiplication of soliton number following each oscillation cycle. Upper-right panels of (e) show the 2D density profiles around the sink, at times marked by the black (i) and gray (ii) dotted lines, respectively. Image size is (25$\mu\mathrm{m})^2$. The lower-right panel in (e) shows line cuts of the corresponding phase profiles. Arrows indicate the phase slips across the solitons within the inner horizon.}  
\label{fig:1}
\end{figure}

Here, we address the role of the Landau instability in a 2D radial flow free from a moving obstacle. We create a particle sink at the center of an otherwise homogeneous atomic superfluid trapped in an optical box. The sink induces fast atom number loss and results in a large inward radial flow, which can be viewed as forming an acoustic black-hole horizon and an inner horizon around the sink. 

We control particle loss rate in the sink through three-body recombination~\cite{weber2003three}, a dissipative process during which three atoms collide to form one bound molecule and one energetic atom, both with kinetic energy large enough to escape a shallow optical trap. Three-body recombination loss scales cubically with the atomic density as $\dot{n} = - L_3 n^3$, where $L_3\approx 4.3\times 10^{-2}~\mu$m$^4/$s is the loss coefficient in our 2D geometry~\cite{SM} and $n$ the 2D density. In our ultracold cesium atomic samples, two-body loss is fully suppressed. We use all conservative potentials in contrast to a related proposal~\cite{sels2020thermal} that utilizes localized one-body loss to generate supersonic flows. 

As illustrated in Figs.~1(a) and~2, a 2D superfluid is initially trapped inside a circular box of potential height $\approx k_\mathrm{B}\times 60~$nK, with a uniform density $n_0 \approx 14~\mu$m$^{-2}$ and a chemical potential $\mu_0 = \hbar^2 n_0 g/m \approx k_\mathrm{B}\times 21~$nK~\cite{SM}. Here, $g\approx 0.42$ is the interaction parameter, $\hbar$ the reduced Planck constant, $m$ the atomic mass, and $k_\mathrm{B}$ the Boltzmann constant. We introduce the sink by ramping on a Gaussian potential of $1/e^{2}$ radius $r_{\rm s}\approx 6.5~\mu$m and depth $V_0\approx k_\mathrm{B}\times 200~$nK at the box center. The attractive potential gives rise to a much higher peak density $> 90\mu$m$^{-2}$ in the sink, leading to more than 250-fold increase in the local three-body loss rate and an estimated total loss rate of $\Gamma = \int_\mathrm{s} |\dot{n}| d^2r \gtrsim 6.5\times 10^{5}$s$^{-1}$ in the sink region. Assuming fluid continuity at $r > r_{\rm s}$, one can estimate the radial velocity as $v(r) = \vec{v}\cdot\hat{r} \approx - \Gamma/[2\pi r n(r)] \lesssim -1~$mm/s, indicating that $v$ can become supersonic outside the sink. 

We first perform theoretical analyses on the stability of this dissipation-induced flow. We model the process through a classical 2D Gross-Pitaevskii equation (GPE) with an additional term accounting for the three-body loss~\cite{SM}. Assuming rotational symmetry, we have found stationary solutions by allowing inflow of atoms at the boundary. We expect such solutions to be close to {\em quasi}-stationary states in a large sample without an inflow. Specifically, for $V_0$ below a critical value $V_{\rm cr}$($\approx k_{\rm B} \times 88~$nK for the chosen parameter values), we find a ground state solution and a transition (`droplet') state that, similarly to the saddle-point solution \cite{little1967decay,langer1967intrinsic} of the Ginzburg-Landau theory, can be interpreted as the fluctuation mediating a phase slip. At $V_0 = V_{\rm cr}$, the solutions merge and disappear through a saddle-node bifurcation, in parallel to the results obtained for conservative flows over obstacles in 1D~\cite{hakim1997nonlinear}. A critical solution for experimentally relevant parameter values is shown in Fig.~1(b). Notice that the ground state develops a small supersonic region when approaching the critical point. We have observed such a correlation also for other parameter values. We therefore interpret disappearance of the static solutions at the critical point as a Landau-type instability. 

Analogies between our results and those of Refs.~\cite{little1967decay,langer1967intrinsic,hakim1997nonlinear} suggest that ramping the potential past the critical value will induce a self-oscillation process, analogous to the soliton train in a conservative 1D flow \cite{hakim1997nonlinear} or a phase-slip center~\cite{skocpol1974phase} in a superconducting wire. This is supported by full numerical integration of a time-dependent 2D GPE~\cite{SM}. An example is shown in the radial plots in Figs.~1(c-f) and the 2D plots in (e). After $V_0$ passes through a critical point, supersonic flow forms; see (d) and (f) for regions with $\Delta = c+v<0$, where $v$ ($c$) is the local flow velocity (sound speed). We evaluate $v=\frac{\hbar}{m}\frac{\partial \phi}{\partial r}$, where $\phi$ is the phase of the wave function. Coincidentally, a train of ring-shaped dark solitons~\cite{kivshar1994ring} (3 clearly visible in this example) are emitted toward the sink center, in time separation $\lesssim 3~$ms. They appear as left-moving dark dips in the radial plots (c) and (e), forcing oscillations in the supersonic flow. We find that this process is insensitive to the potential ramp speed and that soliton emission always accompanies formation of a supersonic flow near the critical point.

Once initiated, a ring soliton's radial motion becomes part of a multiplication process, a remarkable effect absent in 1D black-hole lasers. We point out that a shrinking ring dark soliton cannot stop at the `singularity' at $r=0$~\cite{kivshar1994ring,tamura2023observation}. A soliton first passes through the inner horizon, reaches an inner turning point at $r\geq 0$, and then expands radially back towards the supersonic region (outward-moving dips 
in the density in Fig.~1(e), showing opposite phase slips compared with the inward-moving ones in the phase plot). Upon reaching the inner horizon, the soliton is reflected towards
$r =0$, leaving behind a wake of out-going radiation. During this process, new solitons are continually emitted inward near the inner horizon and also become trapped at smaller radii.
The system thus behaves like an amplifying `soliton laser' mediating oscillating supersonic flows while emitting sound waves out of the black-hole horizon.

\begin{figure}[t]
\centering
\includegraphics[width=1\columnwidth]{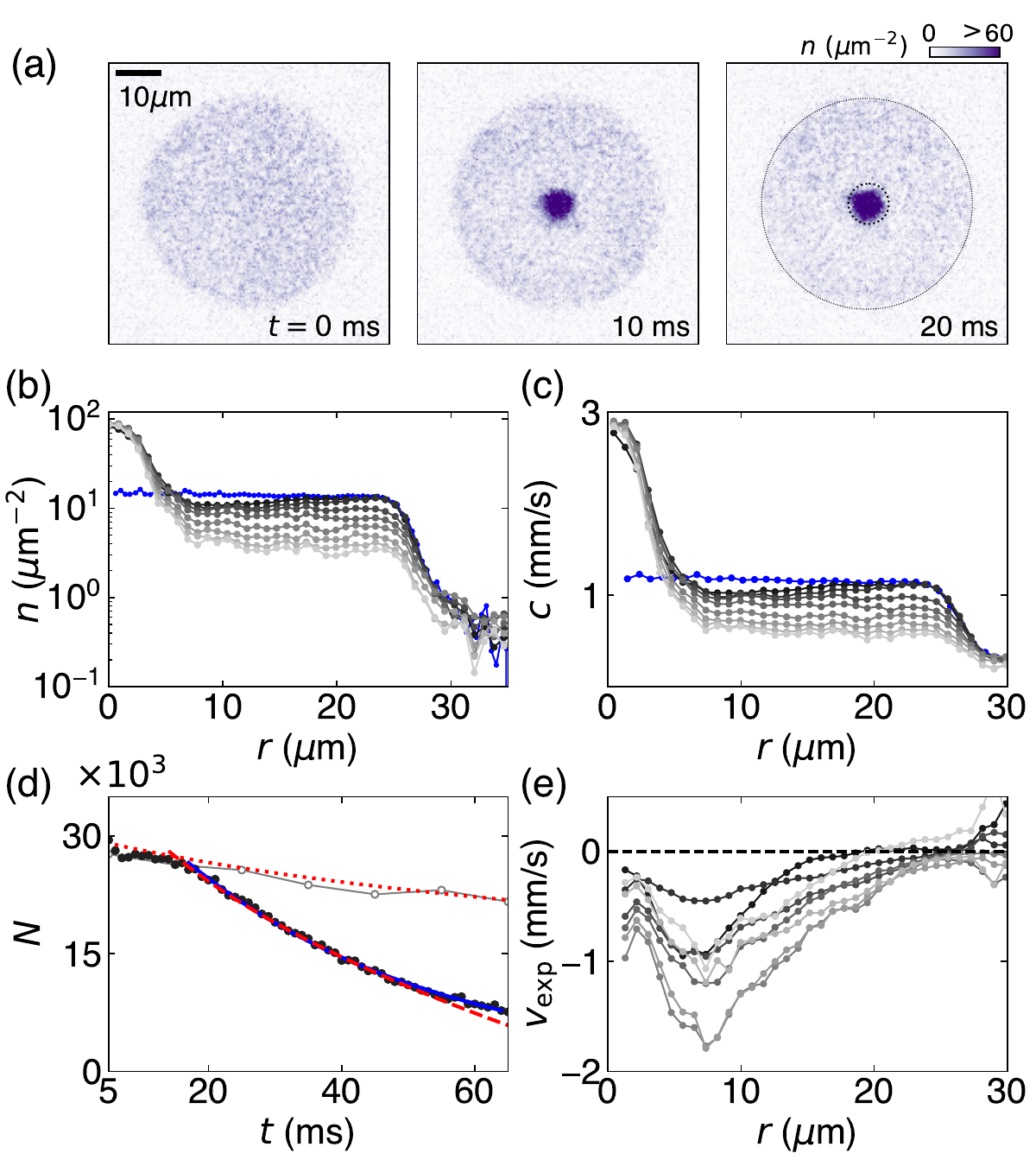}
\caption{Realization of radial supersonic flow in a 2D superfluid. (a) Single-shot in-situ density images measured at the indicated time $t$, before and after the sink is fully ramped on at $t=5$~ms. Dotted (dashed) circles mark $r=26~\mu$m ($5~\mu$m) radius. (b) Radial density profiles $n(r)$ measured at $t=7-63~$ms (dark to light gray circles) with a time interval of 8~ms. Initial density profile (blue circles) is plotted for comparison. (c) Local sound speed $c(r)$ evaluated using profiles in (b). (d) Evolution of integrated atom number $N(t)$ with (filled circles) and without (open circles) the sink, agreeing with a model assuming three-body recombination loss (red curves). Blue dashed curve is a simple exponential fit, giving the total atom number decay rate $\gamma$. (e) Radial flow velocity evaluated using Eq.~(1).}
\label{fig:2}
\end{figure}

In the actual experiment, we have adopted a faster ramp speed ($\Delta t = 5~$ms) to be able to observe the instability before losing many atoms. As shown in the in-situ images in Fig.~2 (a) and averaged radial density plots in (b), shortly after the attractive potential is ramped on, atomic density slightly depletes at $r\lesssim 15~\mu$m outside the sink, showing a strong tendency for the superfluid to flow inwards. As time increases, the density continues to decrease, indicating a continuous flow into the sink region even after the peak density has saturated. Using the radial density profiles, we evaluate the local sound speed $c(r)$, which is nearly uniform and gradually decreases with time to $<1~$mm/s as shown in (c) except within the sink where the density is high. 

In Fig.~2(d), we plot the total atom number $N(t)$, excluding the central region $r \leq r_\mathrm{s}$. Decay of $N(t)$ is consistent with atoms flowing into the sink to compensate for the loss of atoms due to three-body recombination, as described by a simple theory curve (red dashed line) in Fig.~2(d)~\cite{SM}. The overall decay rate $\gamma \approx 27~$s$^{-1}$ is determined by a fit.

Due to finite resolution of our imaging system ($\sim 1~\mu$m), we cannot clearly identify ring dark solitons in situ, as the characteristic width of their density dip is $\xi \approx 1/\sqrt{ng}= 0.2-0.4~\mu$m. We also note that, in a superfluid with preexisting density noise~\cite{tamura2023observation} or imperfect rotational symmetry~\cite{toikka2013snake}, a ring soliton suffers strong snaking instability~\cite{kivshar2000self} and can quickly decay into vortices~\cite{theocharis2003ring} (also $\sim \xi$ wide) that are challenging to measure in situ. This decay has been observed in GPE simulations as well~\footnote{Our simulation is performed under Cartesian grids. In polar grids, ring dark solitons are stable even when they are oscillating in a trap, as shown in Ref.~\cite{toikka2013snake}.}. Generation and decay of ring dark solitons have been reported in our system but with a different experimental setting~\cite{tamura2023observation}.

We can nevertheless identify key signatures of Landau instability and self-oscillation in the measured flow. To extract this information, we compute the local radial flow velocity using the rate of change of total atom number in an annular region bounded by $(r,r_\infty)$ via the expression
\begin{equation}
    v_{\rm exp}(r,t)  =  \frac{1}{2\pi r n(r,t)} \frac{d N(r,t)}{d t},\label{eq:radial_v}
\end{equation}
where $N(r,t) = \int_r^{r_\infty} n(r',t)d^2 r'$ and $r_\infty \approx 40~\mu$m extends well beyond the edge of the box trap. Figure~2(e) plots the radial flow velocity evaluated at various times, showing mostly inward flow $v_{\rm exp}(r)<0$ everywhere for $r\lesssim 26~\mu$m. The magnitude $|v_{\rm exp}(r)|$ increases with decreasing radial position, reaching maximum at around $r\approx 8~\mu$m. It then greatly decreases when approaching the sink region where density becomes high, in qualitative agreement with the flow analyses. Time dependence of the flow, on the other hand, shows an intriguing oscillatory behavior that we now discuss. 

To clearly see the evolution of the superflow, we plot the full spatial-temporal dependence of $\Delta_{\rm exp}(r,t) = c(r,t) + v_{\rm exp}(r,t)$ as shown in Fig.~3(a). Supersonic flow initially appears within a radial interval $(r_\mathrm{in},r_\mathrm{out})\approx (5,10)~\mu$m, enlarging to $\approx(5,15)~\mu$m at later times. This can be viewed as a supersonic flow cavity bounded by a black-hole horizon at $r=r_\mathrm{out}$ and an inner white-hole horizon at $r=r_\mathrm{in}$. For comparison, we also evaluate the flow velocity $v$ obtained in a GPE calculation with the same ramp speed ($\Delta t = 5~$ms) as in the experiment. The result is shown in Fig.~3(b-c).

There is however a striking difference between the experiment and the simulation results. In the experiment, at around $t\approx 10~$ms, shortly after the inward flow becomes supersonic, a sudden change to an apparent outward flow is observed ($v_\mathrm{exp}>0$); see also Fig.~3(c). At larger times, $\Delta_{\rm exp}$ appears to display quasi-periodic short pulses with a primary time period of $t_{\rm p}\approx 4~$ms. This pulsation behavior is from time-dependence of the flow velocity as the sound speed is monotonically decreasing in time. The pulse period is also longer than that of possible collective modes in the sink region, if any are excited. Most surprisingly, these pulses appear to propagate over the entire sample within a small time $\lesssim 2~$ms, which is much shorter than the time period $\gtrsim 20~$ms required for sound waves to traverse the sample.

\begin{figure}[t!]
\centering
\includegraphics[width=1\columnwidth]{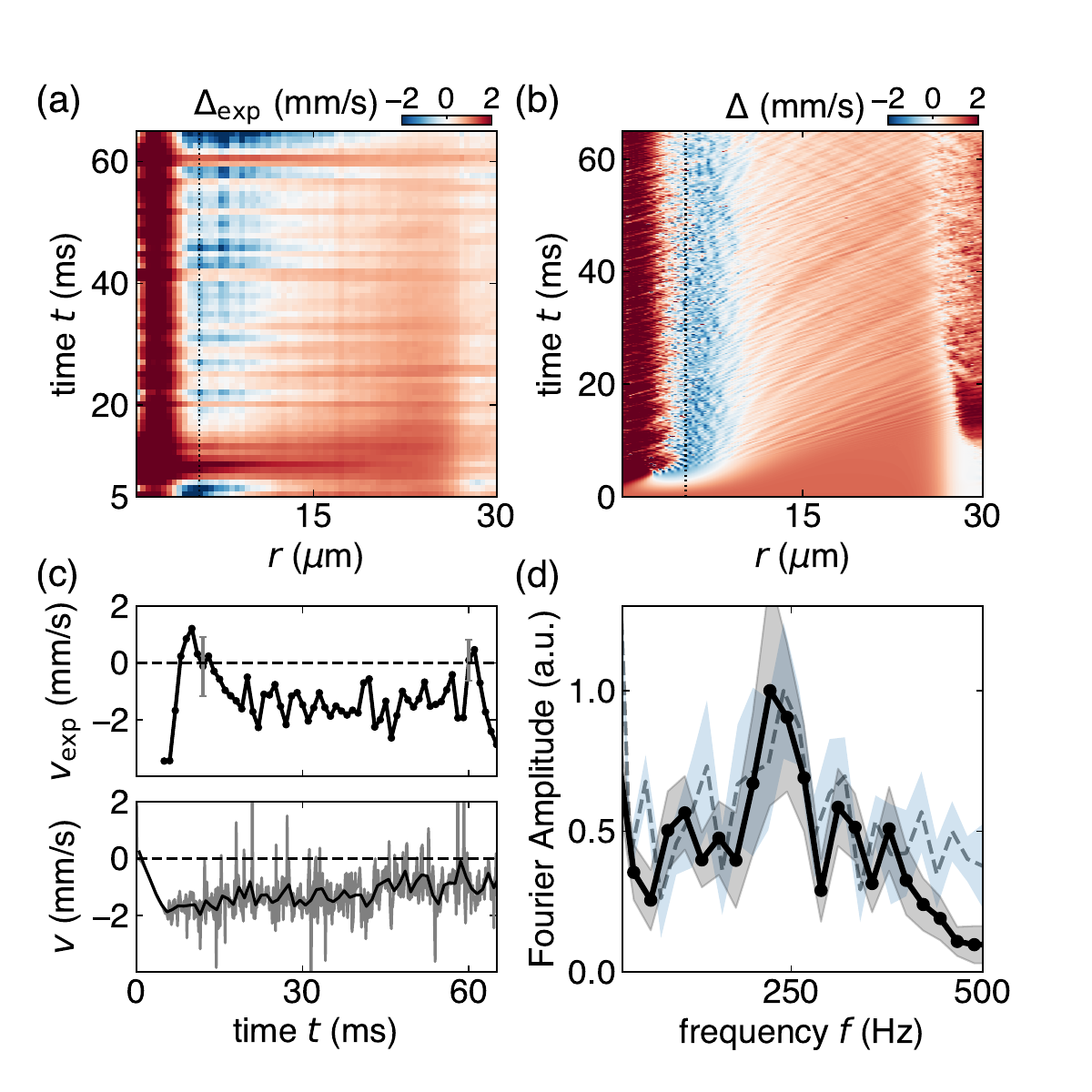}
\caption{Observation of self-oscillating supersonic flow. (a) Measured time evolution (in 1~ms steps) of $\Delta_{\rm exp}(r)$, showing supersonic flow ($\Delta_{\rm exp} <0$) at $r > r_{\rm in}\approx 5~\mu$m and $r < r_{\rm out}$, where $r_{\rm out} \gtrsim 10~\mu$m grows slowly with time. Quasi-periodic pulses of $\Delta_{\rm exp} > 0$ are clearly visible. (b) Calculated $\Delta(r,t)$. (c) Measured (top panel) and calculated (gray curve, bottom panel) flow velocities at positions as shown in the dashed lines in (a) and (b), respectively. A $1~$ms running-average (black curve) is plotted for comparing with experiment. Error bars represent the standard error of the mean. (d) Normalized Fourier spectra of $v_\mathrm{exp}$ (black circles) and $v$ (dashed curve), respectively. Gray and blue shaded bands indicate the corresponding statistical uncertainties.}
\label{fig:3}
\end{figure}

The apparent short pulses could be due to outbursts of atoms from the sink region, traveling at superluminal speeds greater than $1~$cm/s (kinetic energy $>k_\mathrm{B}\times 1~\mu$K). These energetic atoms may come from three-body recombination~\cite{weber2003three,wolf2017state}, with each atom carrying away $2/3$ of the binding energy of the bound molecular state. The closest to the continuum, $6s$ state of Cs$_2$~\cite{mark2007spectroscopy}, has a binding energy $E_{\rm b}\approx k_\mathrm{B}\times20~\mu$K, thus giving an estimated out-going atom velocity of $\approx 4~$cm/s presumably along random directions in 3D. Some of these atoms will be imaged in our apparatus. As the recombination loss occurs primarily in the sink region, we expect it to be modulated by self-oscillations. These effects are not captured in our classical GPE calculations. Another source of energetic atoms, although likely much less prominent, may be the dynamical Casimir effect~\cite{jaskula2012acoustic,hung2013cosmology,eckel2018rapidly,fedichev2004cosmological}, wherein the motion of solitonic defects results in rapid density perturbations in the sink region, possibly capable of exciting short-scale fluctuations ($\lesssim \xi$) with superluminal speeds comparable to $2\pi\hbar/m\xi\sim2$~cm/s.

\begin{figure}[t]
\centering
\includegraphics[width=1\columnwidth]{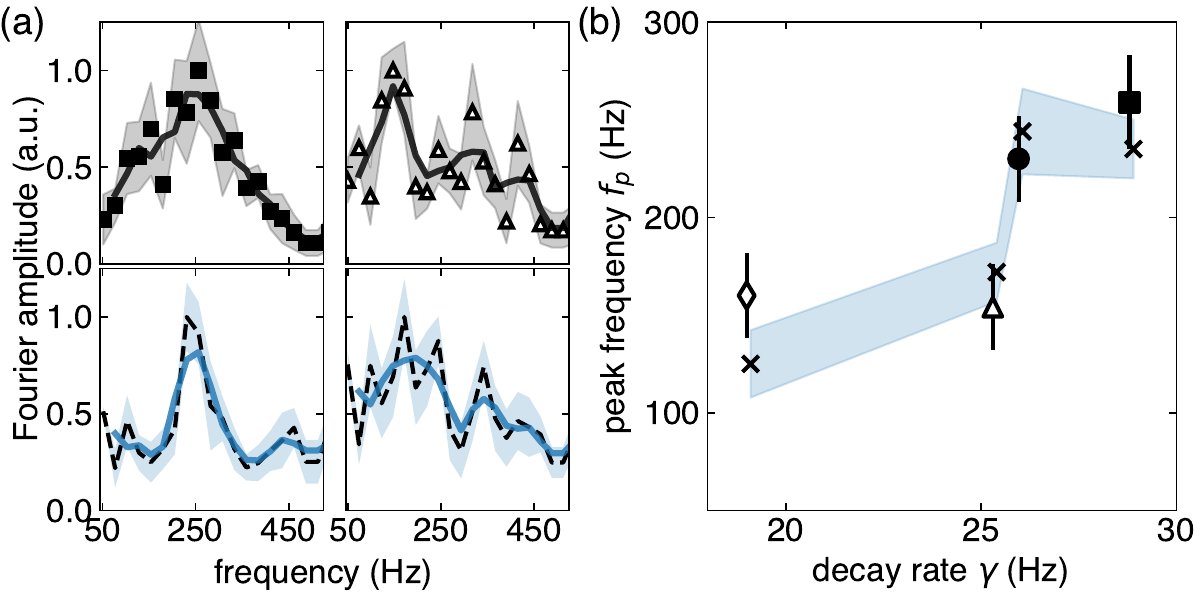}
\caption{Self-oscillation frequencies. (a) Comparisons between the Fourier spectra of $v_\mathrm{exp}$ of two different sink depths, $V_0\approx k_\mathrm{B}\times 200~$nK (filled symbols) and $125~$nK (open symbols), and the spectra of $v$, the flow velocity computed from the GPE wavefunction (dashed curves, bottom panels). Shaded bands indicate the statistical uncertainties. Solid curves are three-point smoothed spectra to guide the eye. (b) Main peak frequencies $f_p$  versus the measured atom number decay rate $\gamma$; Open (filled) symbols identify the values of $V_0$ as in (a).
Crosses mark the peak frequencies identified from GPE calculations. Error bars (shaded band) reflect the uncertainty for experiments (GPE calculations).  In (a) and (b), relevant experimental parameters are $(g, n_0,r_\mathrm{s})\approx (0.42,14,6.5)$ (circles), $(0.48,13, 6.5)$ (squares), $(0.42,25,5)$ (triangles), and $(0.45,20,5)$ (diamonds); the units of length are microns. }
\label{fig:4}
\end{figure}

To further analyze the temporal signature of the fast pulses, we calculate the Fourier spectrum of $v_\mathrm{exp}(t)$. We then compare it with the Fourier spectra of the GPE results averaged over the sink region~\cite{SM} to find possible connection with soliton motions. As shown in Fig.~3(d), the most prominent frequency peak observed in the experiment is at $f_{\rm p} \approx 225~$Hz$~\sim t_\mathrm{p}^{-1}$. This appears to overlap with the main frequency peak in the GPE result. We have verified that $f_{\rm p}$ indeed corresponds to the radial oscillation frequency of tightly trapped solitons, which are reflected upon re-entering the supersonic region.

Experimentally, we have observed the peak frequency to shift slightly to $f_{\rm p} \approx 250~$Hz upon a moderate increase of the atomic interaction; see the upper-left panel in Fig.~4(a). The lower-left panel shows the corresponding frequency peak in the GPE simulation. We have also adopted a shallower depth $V_0\approx k_\mathrm{B}\times 125~$nK and a slightly narrower sink (upper-right panel). The peak frequency appears to reduce to $f_{\rm p} \approx 150~$Hz close to a peak identified in the simulation (lower-right panel). 

In general, the oscillation frequencies should depend on the radius of the horizon, the supersonic flow speed, and the local sound speed. We summarize these results using a single parameter, the total atom number decay rate $\gamma$, which characterizes the overall dissipation rate in the sink. Figure~4(b) summarizes the measurement results, where we find the peak positions showing reasonable agreement with those of GPE calculations. Thus, we believe the short pulses are synchronized with the soliton motion within the horizon. For more frequency dependence on various parameters in GPE calculations, see Supplemental Figures~\cite{SM}. 

In summary, by introducing a stationary, conservative local attractive potential, we observe self-induced supersonic flow initiated solely by fast local three-body recombination. Through observing quasi-periodic emission of superluminal signals, which we attribute to strong nonlinear effects in the sink, we discover that the system develops a self-oscillating dynamics. By measuring the frequency spectrum of the oscillations, we obtain evidence that they are synchronized with the quasi-periodic soliton motion observed in our GPE simulations. Our setup may be considered as a classical analogue of a black-hole laser, with the supersonic region and the `singularity' at the sink center acting as highly reflective cavity mirrors, a pattern distinct from the more familiar 1D black-hole lasers. More generally, our work shows that projecting arbitrary sink potentials to initiate localized dissipation processes may be a valuable way to generate complex flow patterns in atomic superfluids. Our experiments show modulation signal over a period of time, implying that the energy released by three-body recombination continues to be converted to excitations and suggesting that additional instabilities may develop, such as those leading to a turbulent cascade~\cite{khlebnikov2002strong,tsatsos2016quantum,henn2009emergence,kwon2014relaxation,navon2016emergence}. We expect that the momentum distribution and the kinetic energy spectrum of such a cascade~\cite{khlebnikov2002strong,tsatsos2016quantum} can be measured using time-of-flight~\cite{navon2016emergence} and in situ density noise measurements following creation of the supersonic flow. In addition, imaging density defects and counting of solitons and their decay into vortices can be realized in a 2D time-of-flight with increased healing length~\cite{tamura2023observation}. Our work calls for future studies on self-oscillations in dissipative quantum fluids and can potentially find new applications in quantum simulations of curved spacetime~\cite{viermann2022quantum,barcelo2011analogue}.

\begin{acknowledgments}
We thank Chris Greene, Martin Kruczenski, and Qi Zhou for discussions. This work is supported by the W. M. Keck Foundation, the NSF (Grant \# PHY-1848316 and PHY-2409591), and the DOE QuantISED program through the Fermilab Quantum Consortium.
\end{acknowledgments}

\bibliography{BH}

\appendix*
\renewcommand{\thesubsection}{S\arabic{subsection}}
\renewcommand{\figurename}{Fig.}
\renewcommand{\thefigure}{SM\arabic{figure}}
\setcounter{figure}{0}
\renewcommand{\theequation}{S\arabic{equation}}
\setcounter{equation}{0}
\setcounter{page}{1}

\section*{Supplemental Materials}
\subsection{Preparation and detection of a two-dimensional (2D) superfluid}
Our 2D superfluid samples are formed by $N\approx 3\times 10^4$ Bose-condensed cesium atoms loaded into an all repulsive optical box potential with an initial temperature $T < 10~$nK. Detailed descriptions on the condensate formation and trap loading procedures can be found in Refs.~\cite{chen2021observation, tamura2023observation}. The vertical confinement of the optical box offers a trap frequency of $\omega_z \approx 2\pi \times 1.8~$kHz. The atoms occupy the ground state with an harmonic oscillator length of $l_z = \sqrt{\hbar/m\omega_z}\approx 207~$nm, where $\hbar$ is the reduced Planck constant and $m$ is the cesium atomic mass. The horizontal box confinement is provided by a ring-shaped repulsive wall potential with an adjustable radius patterned by a digital mirror device and projected through a high numerical aperture ($\mathrm{NA}\approx 0.6$) objective lens. Cross-section of the wall has an approximate Gaussian form and the width is $5~\mu$m. The potential height is $\approx k_\mathrm{B}\times 60~n$K, where $k_\mathrm{B}$ is the Boltzmann constant. In our quasi-2D geometry, the interaction parameter is $g = \sqrt{8\pi}a_\mathrm{s}/l_z$, where $a_\mathrm{s}$ is the three-dimensional (3D) s-wave scattering length, tunable via a Feshbach resonance~\cite{chin2010feshbach}. 

After box loading, we slowly ramp the scattering length to $a_\mathrm{s}\approx 328 \sim 368~a_0$, where $a_0$ is the Bohr radius, to prepare for the experiments shown in Figs.~\ref{fig:2}-\ref{fig:4} at $g\approx 0.42 \sim 0.48$. We measure the in-situ density profiles by saturated absorption imaging through the high-NA objective. Image resolution ($\lesssim 1~\mu$m) is calibrated using in-situ atomic density noise~\cite{hung2011extracting}. 

To evaluate averaged density and flow velocity profiles, we repeat each experiment condition and perform measurements $\sim50$ times to record in situ density images. To minimize error due to shot-to-shot atom number fluctuations, at each hold time we select images with total atom number typically within $\pm10\%$ deviation from a mean number for analyses. Care has been taken to evaluate the mean without being biased by large fluctuations. We estimate the error of the mean is $\lesssim 1.2\%$. Typically, around $\sim 30$ images are selected to evaluate the average density and flow velocity profiles.

\subsection{Calibration of the attractive potential}
The attractive (sink) potential is formed using a far-off resonant optical beam at $876~$nm wavelength, with an adjustable beam size and projected through the high-NA objective lens. We calibrate the potential depth $V_0$ at a much lower optical power to avoid fast three-body recombination loss. The depth is equal to the local chemical potential difference $\Delta \mu = \hbar^2 g\Delta n/m $ between the sink region and the surrounding superfluid region, where $\Delta n$ is the density difference. We calibrate the width $r_\mathrm{s}$ of the Gaussian beam by measuring the size of the density peak. For a strong potential, we compare time-dependent GPE calculations with flow measurement results, through adjusting the width and depth of the sink potential by $10\% \sim 20\%$ around the calibrated value, and find good agreement. In the main text, we report the adjusted values of the sink potential.

\subsection{Three-body recombination loss rate}
We describe three-body recombination loss by the rate equation $\dot{n} = - L_3 n^3$, with a cubic dependence on the 2D density $n$. In our quasi-2D geometry, the vertical oscillator length $l_z \gg a_\mathrm{s}$ is much bigger than the 3D scattering length and the three-body recombination process remains 3D in nature. We relate the rate coefficient $L_3$ in quasi-2D with the 3D loss coefficient $l_3$ by integrating out the ground state density distribution along the $z$-axis. We have $L_3 = l_3/(\pi l_z^2\sqrt{3})$, where $l_3 = C \frac{\hbar}{m} a_\mathrm{s}^4 \approx 0.356 \times\frac{\hbar}{m} (l_z g)^4$. $C\approx 225$ has been measured in Ref.~\cite{weber2003three} using cesium thermal gas. In the experiment, we have tuned to a larger scattering length to increase the three-body loss coefficient to $L_3 \approx 0.042\sim 0.071~\mu{\rm m}^4$/s. We note that our measured particle loss rate from the sink is consistent with a loss coefficient of thermal gas without the $\frac{1}{3!}$ suppression factor for identical bosons.

In Fig.~2(d), we model the time dependence of the total particle number $N$ outside the sink region ($N=\int_{r>r_\mathrm{s} }n d^2r$) by 
\begin{equation}
\frac{dN}{dt} = -L_3 \int n^3 d^2r\,,\label{eqSM:lossN}    
\end{equation}
where $n=n_\mathrm{s} + n_\mathrm{bg}$, $n_\mathrm{s}(r) = n_\mathrm{p} e^{-2r^2/\sigma^2}$ is a saturated Gaussian density distribution at the sink, $\sigma$ the radius, $n_\mathrm{p}$ the peak density, $n_\mathrm{bg}=N/(\pi R^2)$ the background uniform density, and $R$ the sample radius. We determine $(n_p, \sigma) \approx (80~\mu{\rm m}^{-2}, 4.8~\mu{\rm m})$ by fitting the sink density distribution. Time dependence of $N(t)$ can be completely determined by Eq.~(\ref{eqSM:lossN}). In Fig.~2(d), we plot the dashed curve from $t_i=14~$ms with $n_\mathrm{bg}(t_i)=12.5~\mu{\rm m}^{-2}$ and $R= 27~\mu$m, and plot the dotted curve without the sink by setting $n_\mathrm{p}=0$ and $t_i=0$.

\subsection{Stationary solution for the supersonic flow induced by a particle sink}

We look for stationary solutions to the classical 2D Gross-Pitaevskii equation (GPE) 
with an additional term accounting for three-body recombination:
\be
i \hbar \partial_t \psi = \frac{\hbar^2}{2m} (- \nabla^2 + 2 g n - i \gamma_3 n^2 ) \psi 
+  [V(r) - \mu ] \psi \, .
\label{eqSM:GP}
\ee
Here $\psi$ is the complex amplitude, $n = \psi^\dagger \psi$ is the atom density, $\gamma_3 = m L_3/\hbar$, and
$V(r)$ is the sink potential, which we take in the Gaussian form
\[
V(r) = - V_0 \exp(-2 r^2 / r_\mathrm{s}^2) \, .
\]
At the classical level, it is consistent to assume perfect rotational symmetry, so 
$\psi \equiv \psi(r,t)$, and $\nabla^2$ is the radial part of the 2D Laplacian.
The radial coordinate $r$ takes values in the range $0 \leq r \leq R$. At $r = 0$, we impose the standard regularity condition $\partial_r \psi = 0$. Because the system is dissipative (with a
3-body loss coefficient $\gamma_3$), a stationary solution can only exist if we allow inflow of atoms
at the boundary. To achieve that, we impose different boundary conditions for the real
and imaginary parts of $\psi$ at $r=R$---Neumann for the real part and Dirichlet for the imaginary:
$\partial_r \mbox{Re} [\psi] = 0$, $\mbox{Im} [\psi] = 0$. 
These permit a nonzero inflow velocity at $r=R$, which however is not specified a priori but 
will be found as a part of the solution.
The idea is that, if $R$ is sufficiently large, such a stationary solution (if any is found) will 
be close enough to a {\em quasi}-stationary state that would be obtained in a large sample without 
any inflow at the boundary.

To look for stationary solutions, we set the left-hand side of Eq.~(\ref{eqSM:GP}) to zero, discretize the
right-hand side on a uniform grid of $N$ points (typically $N=500$) and solve the resulting system
of equations numerically by the multidimensional Newton-Raphson (NR) method. We define a real $2N\times 2N$ Jacobian matrix with elements $J_{ik} = \partial F_i / \partial u_k$, where $\{ F_i \}$ is the set of the real and imaginary parts of the right-hand side of Eq.~(\ref{eqSM:GP}) at the grid points, and $\{ u_k \}$ is the set of
those of $\psi$. It is convenient to vary
the potential depth $V_0$ while keeping all the other parameters fixed. For $V_0$ below a certain critical value $V_{\rm cr}$, we find two solutions. For either, eigenvalues of the Jacobian are all 
real and, for one of the solutions, are in fact all positive. The other solution has one negative mode.
We refer to the first solution as the ground state, and to the second as the droplet, as discussed in the main text. At $V_0 = V_{\rm cr}$ the solutions merge and disappear through a saddle-node bifurcation. A critical solution is shown in Fig.~1(b).

\subsection{Time-dependent GPE simulation}
We perform time-dependent simulations of Eq.~(\ref{eqSM:GP}) using code described in \cite{antoine2014gpelab,antoine2015gpelab} for matching experimental conditions. The initial ground state wave function is obtained by first setting $\gamma_3=0$ and $V_0=0$, while adding a confining potential of the form
\begin{equation}
    U(r) = \frac{U_0}{2} \left[1+\mathrm{erf}\left(\frac{r-R}{w}\right)\right] \,.
\end{equation}
Here $U_0=k_\mathrm{B}\times 60~$nK is the trap strength, $w=2.5~\mu$m, and $\mathrm{erf}(r)$ is the error function that sets $U(r\lesssim R)=0$ and $U(r\gtrsim R)=U_0$ for us to obtain a localized initial stationary solution. Physical simulation box size is $70\times70~\mu$m$^{2}$ with typically $257\times257$ or $513\times513$ grid points. 

To simulate the dynamics, we set $\gamma_3$ in Eq.~(\ref{eqSM:GP}) to the experimental value and evolve the time-dependent GPE with a linear ramp of attractive potential to a final depth of $V_0 = k_\mathrm{B} \times 200~$nK (or $V_0 = k_\mathrm{B} \times 125~$nK). The simulation time step is $1~\mu$s and the time duration of ramp $\Delta t$ is indicated in the main text. 

\begin{figure}
\centering
\includegraphics[width=0.9\columnwidth]{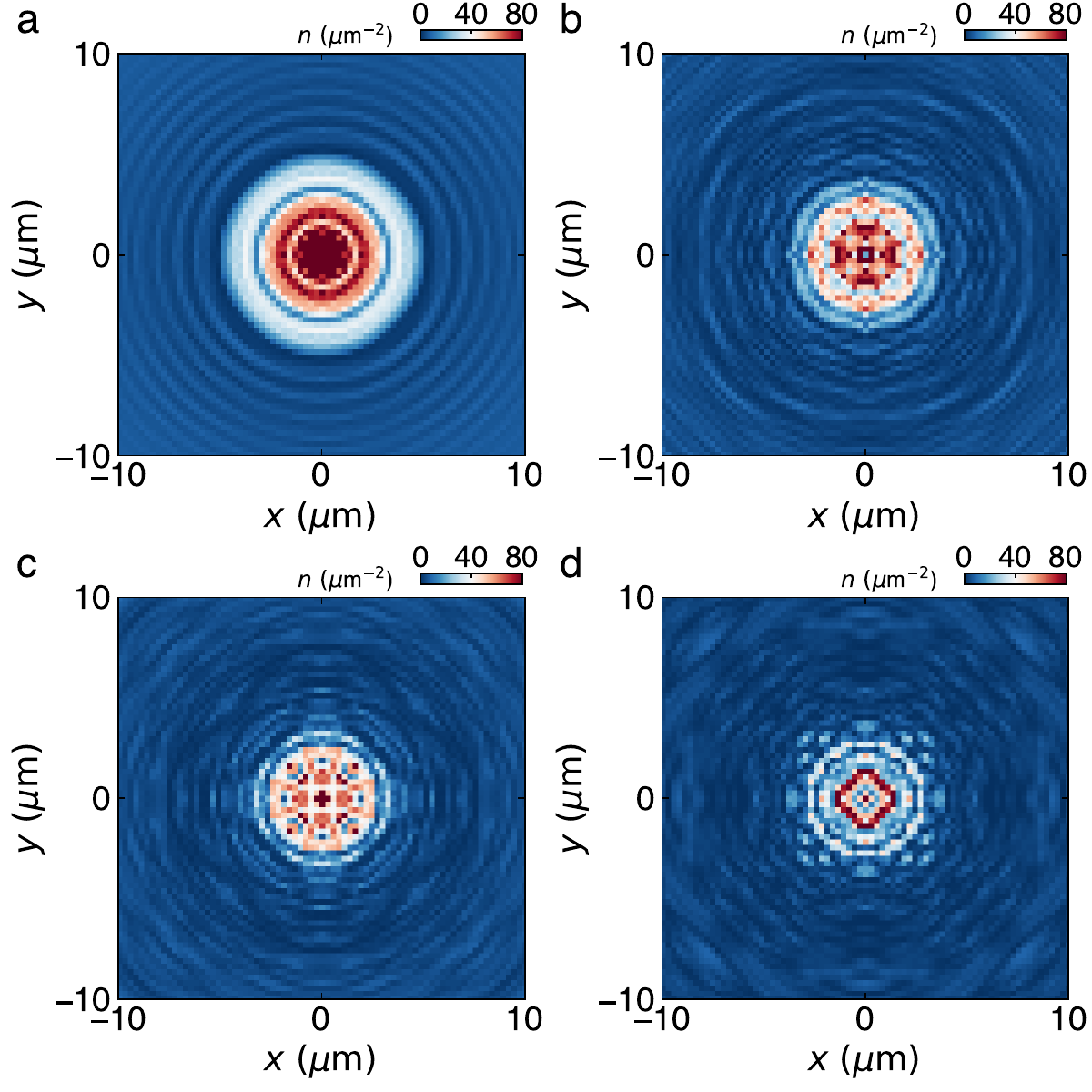}
\caption{2D density profiles around the sink from the GPE simulation as shown in Fig.~\ref{fig:1}(e). Plotted times are (a) 43~ms, (b) 56~ms, (c) 62~ms, and (d) 79~ms, respectively. Image size is ($20~\mu$m)$^2$.} 
\label{extfig:RDS}
\end{figure}

\begin{figure}
\centering
\includegraphics[width=1\columnwidth]{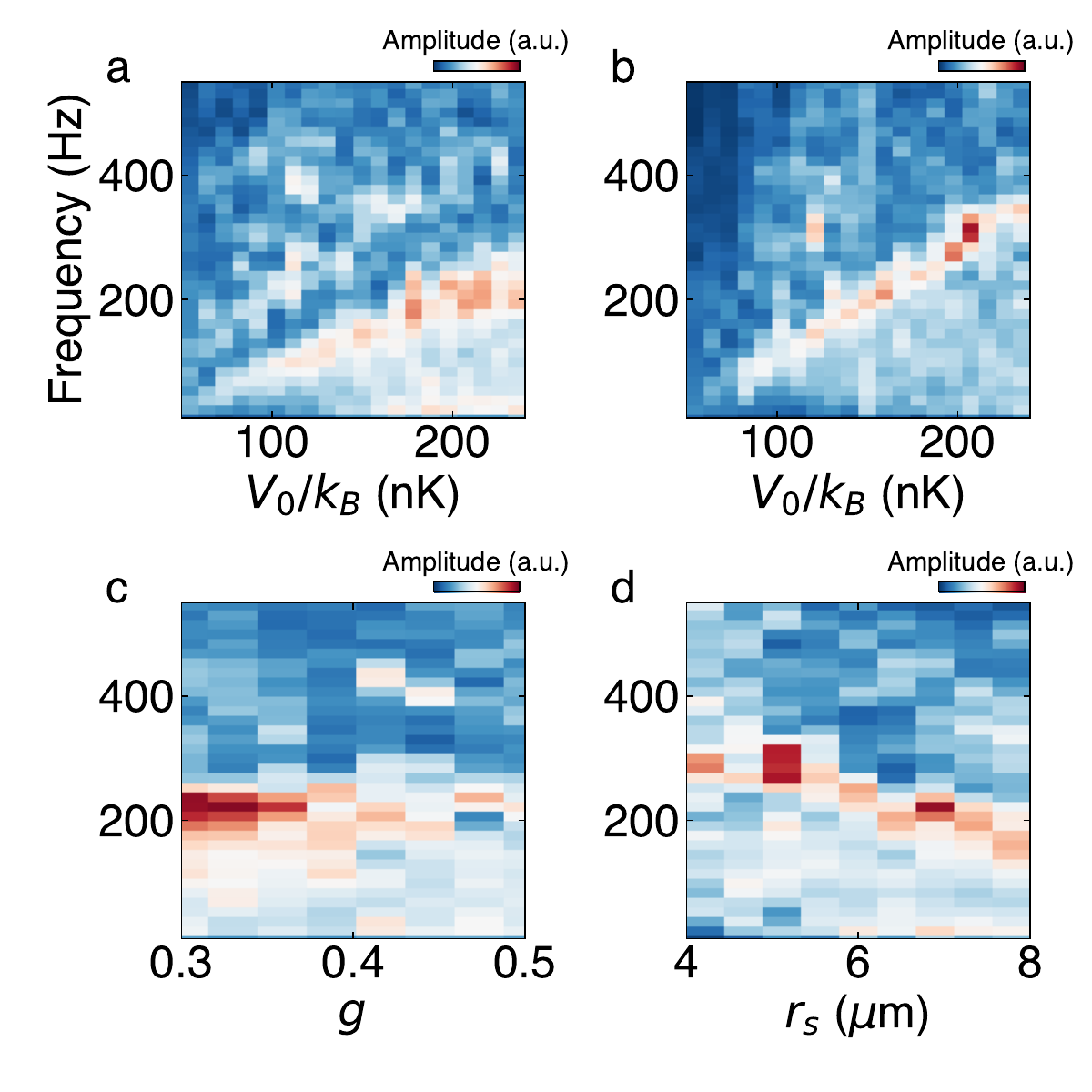}
\caption{Dependence of the Fourier spectra of GPE simulations on various parameters. (a-b), Fourier spectra of calculated flow $v$ versus depth $V_0$ for fixed interaction parameter $g=0.42$ and attractive Gaussian beam radius $r_s$= 7$\mu$m (a), and 5$\mu$m (b), respectively. (c) Fourier spectra of calculated flow $v$ versus interaction parameter $g$ for fixed $V_0/k_B=200~$nK and $r_s=6.5~\mu$m. (d), Fourier spectra of calculated flow $v$ versus $r_s$ for fixed $V_0/k_B=200~$nK and $g=0.42$. In all the numerical simulations, the attractive potential is ramped on in 5~ms, as in the experiment, and the initial total atom number is set to $N=3\times 10^4$.} 
\label{extfig:1}
\end{figure}

\begin{figure}
\centering
\includegraphics[width=1\columnwidth]{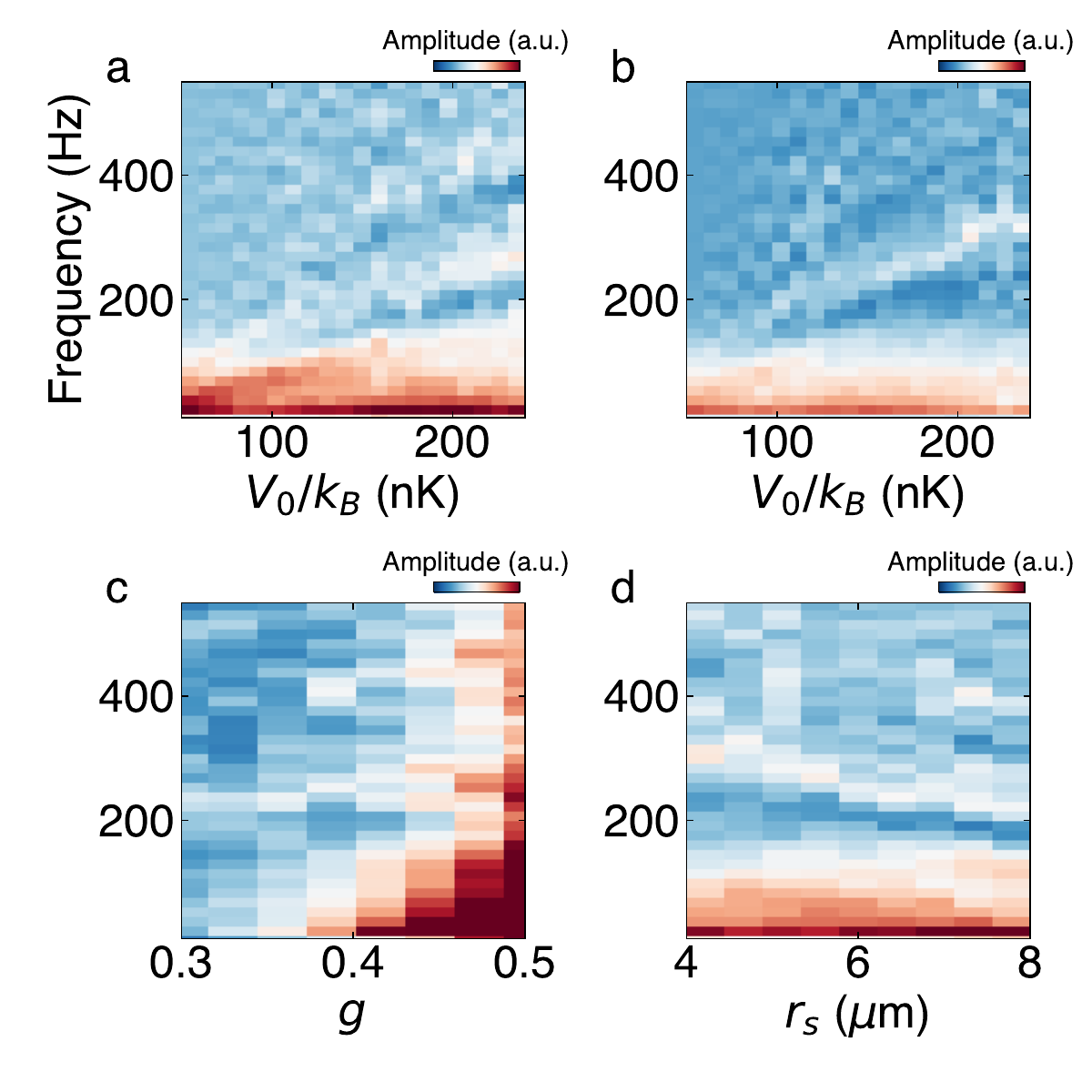}
\caption{Dependence of the Fourier spectra of the flow velocity $v_{\rm GPE}$, evaluated using Eq. (1) in the main text, on the parameters for the same GPE simulations as in Fig.~\ref{extfig:1}.} 
\label{extfig:2}
\end{figure}

We observe trains of ring dark solitons (RDSs) emitted in cases of fast ($\Delta t =5~$ms) and slower ramps ($\Delta t >10~$ms). Persistent self-oscillations of solitons are observed once the potential depth passes through a critical value $V_\mathrm{cr}$ close to the prediction by the stationary state solutions. More simulation results can be found in Figs.~\ref{extfig:RDS}, ~\ref{extfig:1}, and ~\ref{extfig:2}. 

We note that RDSs suffer from snaking instability even in GPE calculations in the absence of preexisting perturbations. Figure~\ref{extfig:RDS} illustrates sample late time images from the GPE simulation as shown in Fig.~\ref{fig:1}(e), showing distorted RDSs. The instability can be triggered by numerical variations (deterministic error) across the simulation grids that break rotational symmetry~\cite{theocharis2003ring,toikka2013snake}. This is particularly severe when an RDS shrinks to a small radius comparable to the grid size. Eventually, an RDS will decay into a ring of vortex dipoles. While this is observed in our 2D GPE simulations, the impact of this instability on soliton self-oscillating dynamics requires further investigations. 


\subsection{Evaluating the Fourier spectra}
To test if the pulses in measured supersonic flows are synchronized with soliton oscillations, we compare the Fourier spectra of $v_\mathrm{exp}(t)$ obtained in experiment and $v_\mathrm{GPE}(t)$, evaluated using Eq.~(\ref{eq:radial_v}), as well as $v(t)$, obtained using the probability current density, in GPE simulations. Due to finite experiment time and resolution (total time $t_\mathrm{max}=55\sim65~$ms with 1~ms time step), our measured Fourier spectrum has a limited frequency resolution $\sim t_\mathrm{max}^{-1}\approx 20~$Hz and sensitivity up to $f=500$~Hz. We apply a boxcar time-window between $t=15\sim65~$ms with a $\pm 5$~ms variable length to evaluate an averaged Fourier spectrum as shown in Fig.~\ref{fig:3}(d) and Fig.~\ref{fig:4}(a). To evaluate the Fourier spectra based on GPE simulations, we evaluate a 1~ms running average of $v(t)$ and $v_\mathrm{GPE}(t)$, respectively, as shown in Fig.~\ref{fig:3}(c), and then calculate the Fourier spectra in the same window-averaging procedure as described above. Since we are testing possible synchronization with soliton motion, we average the GPE spectra over a range of radial positions within the sink. In Fig.~\ref{fig:3}(d) and the lower-left panel of Fig.~\ref{fig:4}(a) ($V_0/k_\mathrm{B}=200~$nK and $r_\mathrm{s}=6.5~\mu$m), we show the spectra averaged over $r\leq5~\mu$m as these spectra clearly exhibit prominent frequency peaks and agree well with our experiment result. For the lower-right panel of Fig.~\ref{fig:4}(a) ($V_0/k_\mathrm{B}=125~$nK and $r_\mathrm{s}=5~\mu$m), we average over $r\leq3.5~\mu$m. 

Compared to the spectra of $v$ in Fig.~\ref{extfig:1}, the main oscillation peaks of $v_{\rm GPE}$ in Fig.~\ref{extfig:2} appear weaker due to contamination by a low frequency background signal likely caused by slow dynamics in the density profile. Thus, we show only the Fourier spectra of $v$ in the main text. In Figs.~\ref{extfig:1} and \ref{extfig:2}, weak multi-tones other than the main peak in the Fourier spectra are attributed to anharmonicity of the oscillation signal.


\subsection{Likelihood analysis considering finite signal-to-noise ratio}
We evaluate the flow speed based on Eq.~(\ref{eq:radial_v}), using the rate of change of the total atom number. Realistically, fluctuations in the total atom number could contribute partially to burst-like signals as seen in Fig.~\ref{fig:3}(c), where the flow speed $v_\mathrm{exp}$ at all $r>0$ simultaneously rises or falls. However, this type of noise would appear randomly in time and would not reveal a specific oscillation frequency.

\begin{figure}
\centering
\includegraphics[width=1\columnwidth]{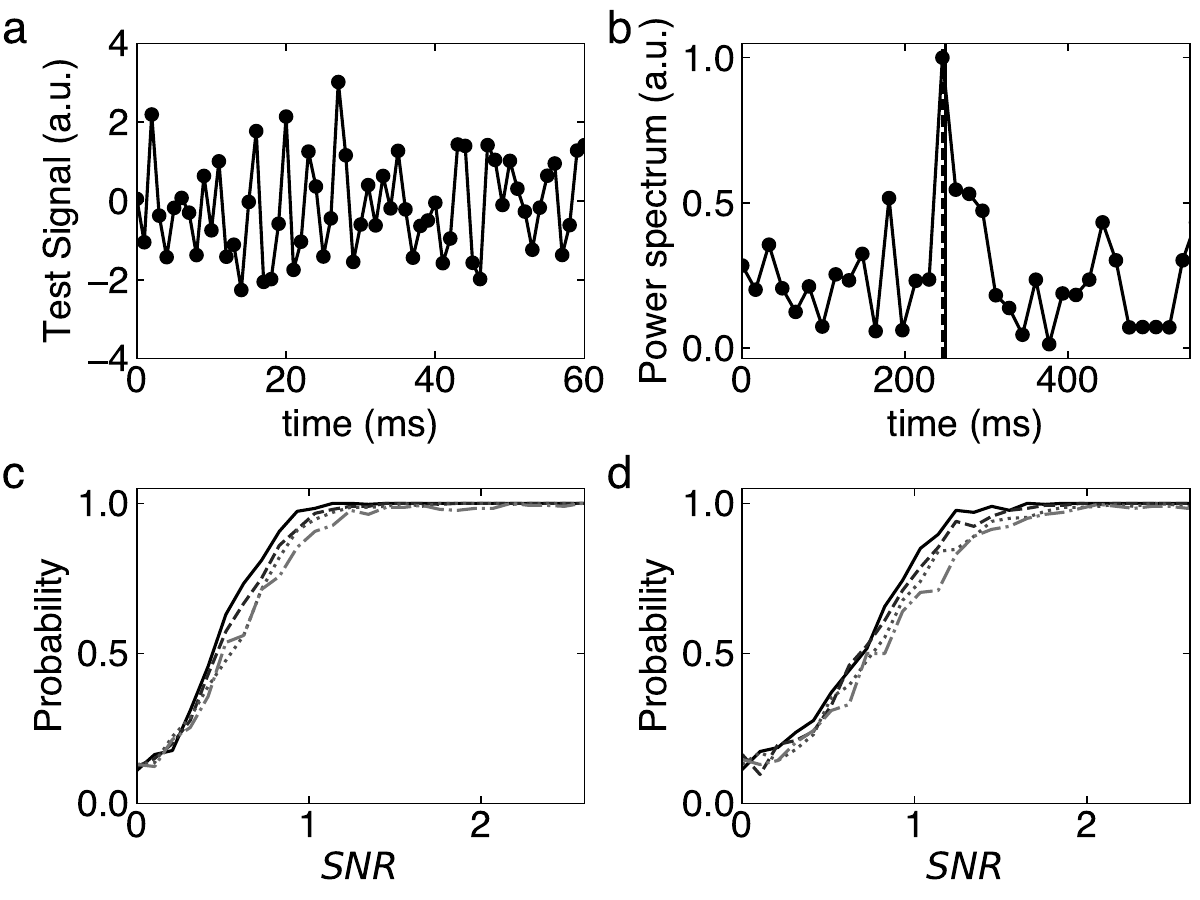}
\caption{(a) Sample numerical sine wave ($f=250~$Hz) of signal-to-noise ratio $\mathrm{SNR}=1$ and with four random phase jumps. (b) The corresponding Fourier spectrum. Vertical dashed line marks the peak frequency $f_p$. Solid line marks true signal frequency. (c,d) Probability for observing the peak frequency within the window $f_p = 250~$Hz$\pm 2\Delta f$ under different SNR, using (c) a sine wave or (d) a square wave as the test waveform. Here, $\Delta f=1/(60~\mathrm{ms})\approx 17~$Hz is the resolution. The phase of the waveform is allowed to randomly jump for one (solid), two (dashed), three (dotted), and four (dash-dotted) times, respectively. } 
\label{extfig:3}
\end{figure}

To quantitatively investigate the likelihood that our measured oscillation spectrum results purely from noise, we consider Gaussian noise in $t_\mathrm{max}=60$~ms samples with 1~ms steps to simulate the effect of atom number fluctuations. We then superimpose the noise with signals of the form $v(2\pi f t+\phi)$, where $v(x)$ assumes either a sinusoidal waveform or a square wave, $f=250~$Hz, and $\phi$ is a random phase. We allow $\phi$ to randomly change multiple times to simulate possible asynchronous oscillation signals. We vary the amplitude of $v$ to obtain test samples of different signal-to-noise ratio (SNR). Figure~\ref{extfig:3}(a) shows a test sample of $\mathrm{SNR}=1$. We compute the Fourier spectrum and identify the maximum peak frequency $f_p$, as shown in Fig.~\ref{extfig:3}(b). Figure~\ref{extfig:3}(c) and (d) show the probability of the identified peak frequency $f_p$ to locate within $250$~Hz $ \pm 2\Delta f$, where $\Delta f=t_\mathrm{max}^{-1}\approx 17~$Hz is the resolution. Each point in Fig.~\ref{extfig:3}(c,d) is evaluated using 300 samples.

As shown in Fig.~\ref{extfig:3}(c,d), the likelihood that $f_p\approx 250~$Hz results purely from noise ($\mathrm{SNR}\approx0$) is around $15\%$. This means that only one out of seven complete experiments like Fig.~\ref{fig:3} could show a false positive signal. For multiple experimental datasets in this work, there is then very little accumulated probability that random noise could produce the observed peaks near the GPE results. On the other hand, for $\mathrm{SNR}\approx 1$ (closer to the quality of our presented data), the likelihood of correctly observing a peak approaches unity ($\gtrsim95\%$ for sine waves and $\gtrsim 80\%$ for square waves, respectively). This suggests that the bursts are much more likely to result from the actual oscillating signals rather than from noise. 

\end{document}